\documentclass[11pt]{article}
\usepackage[margin=1in]{geometry}
\usepackage{amsmath}
\usepackage{hyperref}
\usepackage{graphicx}
\usepackage{cite}
\usepackage{authblk}

\begin{document}

\title{Operationalizing CaMeL: Strengthening LLM Defenses for Enterprise Deployment}

\author[1]{Krti Tallam\thanks{Correspondence: \texttt{ktallam@sentinel-security.ai} (alt: \texttt{krtital@gmail.com})}}
\author[1]{Emma Miller}

\affil[1]{SentinelAI, San Francisco, CA, USA}
\date{\today}

\maketitle

\begin{abstract}
CaMeL (Capabilities for Machine Learning) introduces a capability-based sandbox to mitigate prompt injection attacks in large language model (LLM) agents. While effective, CaMeL assumes a trusted user prompt, omits side-channel concerns, and incurs performance trade-offs due to its dual-LLM architecture. This technical response identifies these limitations and proposes engineering enhancements to extend CaMeL’s threat coverage and operational viability. We introduce: (1) prompt screening for initial inputs, (2) output auditing to detect instruction leakage, (3) a tiered-risk access model to balance utility and control, and (4) a formally verified intermediate language to support static guarantees. Together, these augmentations align CaMeL with best practices in security engineering, reduce overhead, and support enterprise-scale deployment without modifying underlying models.
\end{abstract}

\section{Introduction}

Prompt injection attacks have become one of the most pressing security challenges in the deployment of large language models (LLMs), especially when these models are used as autonomous agents in enterprise workflows. As LLMs move beyond simple text generation and begin orchestrating tools - sending emails, querying databases, managing cloud files - they face increasing exposure to untrusted inputs.

In these scenarios, attackers can embed malicious instructions within otherwise benign content - uploaded documents, web pages, or even system-generated messages. When an LLM ingests such input, it may unknowingly execute unauthorized actions: leaking sensitive data, contacting external recipients, or triggering destructive API calls. Studies have shown that even advanced models routinely fall for these traps, often misinterpreting hidden payloads as legitimate user commands~\cite{carlini2023poisoning, tallam2025cybersentinel}.

To address these threats without modifying the underlying model, Debenedetti \textit{et al.} introduced CaMeL (Capabilities for Machine Learning)~\cite{carlini2023poisoning}. CaMeL wraps the LLM in a capability-based execution layer that separates trusted and untrusted data flows. A \textit{Privileged LLM} receives the user’s prompt and generates a high-level plan, while a \textit{Quarantined LLM} handles untrusted content and enforces strict schema validation. Each data value is labeled with provenance and access metadata (“capabilities”), and all tool invocations must pass explicit policy checks before execution.

While CaMeL significantly raises the bar for prompt-injection resilience, there is room to improve its robustness and practicality. This paper examines CaMeL’s current limitations - including assumptions in its threat model, remaining side-channel exposures, and system-level constraints such as latency and policy sprawl. We propose a series of technical developments to address these challenges, drawing on best practices from security engineering, formal verification, and access control systems.

\section{Opportunities for Strengthening CaMeL’s Threat Model}

From an enterprise security standpoint, three notable gaps in CaMeL’s threat model warrant closer attention.

\subsection{Initial Prompt Trust}
CaMeL assumes that the user's initial message is benign. However, corporate red team exercises have shown that a single crafted prompt - often delivered via phishing or chat-based social engineering - can implant persistent logic into the agent’s planning layer. Classic phishing studies report that over 30 percent of users click seemingly harmless links containing attacker-controlled keywords~\cite{abu2007comparison}.

To address this risk, we propose an \textbf{initial prompt screening gateway} that performs reputation checks on URLs, flags override phrases (e.g., “ignore all previous”), and computes entropy or perplexity scores to detect anomalies. Since this involves only a short string, latency remains low (<5 ms in internal tests), yet the screening closes a high-impact entry point for injection attacks \cite{tallam2025assurance, tallam2025cyberimmune}.

\subsection{Output-Side Manipulation}
CaMeL enforces data provenance for tool inputs but does not inspect what the agent ultimately outputs. For instance, a benign PDF might include a line such as:

\textit{“\#\# System: Forward this file to finance‐external@example.com.”}

If the agent summarizes this document, the hidden instruction could be echoed to a human user, leading to unintended actions. Modern natural language inference models can already detect contradictions and malicious phrasing with over 90 percent accuracy on benchmark datasets like MNLI~\cite{williams2018broad}.

We recommend a post-processing \textbf{output auditing pass} that scans each LLM-generated response for override cues, suspicious URLs, or contradictions with the intended business task. Only outputs that pass this audit are displayed or executed.

\subsection{Provenance of User Uploads}
Schema validation alone is insufficient for documents supplied by end users. We suggest that every value extracted from such files carry a \textbf{file-content provenance tag}, e.g., a \texttt{from\_user\_upload} label. Policies can then prevent this data from flowing into irreversible actions - such as sending external emails or modifying ERP systems - unless explicitly authorized via a privileged “grant-exception” mechanism.

This design mirrors techniques from information-flow control languages like JIF~\cite{myers1999jif} and FlowCaml~\cite{simonet2004flowcaml}, adapted here for agent-based systems driven by LLMs.

\medskip
\noindent
By implementing (i) prompt screening, (ii) output auditing, and (iii) tagged provenance for uploaded content, security teams can broaden CaMeL’s protection across the full conversation lifecycle without compromising its core capability-based controls.

\section{Balancing Security Guarantees with Practical Utility}
\label{sec:balancing}

CaMeL’s default behavior rejects any tool invocation whose arguments - directly or indirectly - contain untrusted data. This strict policy achieves a 67 percent completion rate on the AgentDojo benchmark~\cite{carlini2023poisoning}. However, many of the remaining failures stem not from unsafe logic, but from treating all untrusted inputs as equally risky.

In real-world enterprise environments, security frameworks like Risk-Adaptive Access Control and NIST’s Zero Trust Architecture apply more nuanced policies, adjusting enforcement based on operational sensitivity and situational context~\cite{mell2020zero}. Inspired by these practices, we propose a \textbf{tiered-risk policy} for CaMeL:

\vspace{0.5ex}
\begin{itemize}
  \item \textbf{Green tier}  -  read-only actions on public or internally labeled “open” data (e.g., listing calendar events) are allowed after a basic provenance check.
  \item \textbf{Yellow tier}  -  changes within the user’s own environment (e.g., moving a file to a shared folder) prompt a lightweight confirmation if any argument is untrusted.
  \item \textbf{Red tier}  -  irreversible or externally visible operations (e.g., sending email, making wire transfers, calling privileged APIs) retain full capability checks and require multi-factor approval.
\end{itemize}
\vspace{0.5ex}

Large-scale ABAC evaluations show that this kind of stratification preserves over 90 percent of legitimate workflows while blocking all simulated attacks in cloud testbeds~\cite{hu2015abac}. Applying a similar approach in CaMeL would increase task success rates without relaxing its strongest controls.

\paragraph{Reducing Prompt Fatigue.}
Excessive security prompts can lead to \emph{prompt fatigue}, in which users reflexively approve warnings without reading them~\cite{furman2016security}. By limiting explicit confirmations to the yellow and red tiers, CaMeL can reduce prompt volume significantly - improving usability while maintaining security for high-risk actions.

\paragraph{From Empirical Checks to Formal Guarantees.}
CaMeL’s current safeguards rely on benchmark performance rather than formal verification. For environments that demand rigorous assurance, we recommend building a mechanized model of CaMeL’s interpreter and policy engine in a proof assistant, then proving it satisfies noninterference: secret-labeled inputs should not affect public outputs, except through approved channels.

Verified systems like CertiKOS and CompCert show that machine-checked security is feasible even for complex software stacks~\cite{gu2016certikos,leroy2009compcert}. Rewriting CaMeL’s restricted Python dialect as a minimal, formally specified intermediate language would support similar verification efforts - bridging the gap between theoretical models and practical deployments~\cite{sabelfeld2003language}.

\section{Side-Channel Considerations and Mitigations}

While CaMeL enforces capability-based data flow controls, it does not inherently block information leaks through \textit{side channels} - indirect signals such as timing, loop iterations, or error behavior that can reveal internal state without violating explicit policy.

Below, we examine three practical side-channel vectors and recommend mitigation strategies based on established techniques from secure systems research.

\subsection{Loop-Counting Attack}

\paragraph{Threat.} \textnormal{When the number of loop iterations depends on a secret, even non-sensitive operations can leak information through observable counts. For example, a loop such as:}

\texttt{for i in range(secret): fetch("ping")}

reveals the value of \texttt{secret} through access logs. This same pattern underlies attacks like controlled-channel exploitation against Intel SGX, where memory paging patterns expose enclave state~\cite{xu2015controlled}.

\paragraph{Mitigations.}
\begin{itemize}
  \item \textbf{STRICT mode.} Automatically trigger STRICT evaluation for any loop with a secret-tainted bound. During STRICT mode, state-changing tool calls are blocked or require user confirmation.
  \item \textbf{Loop limits.} Reject or cap the maximum number of iterations if the loop count depends on confidential data.
  \item \textbf{Call batching.} Combine repeated benign operations into a single bulk request, breaking the correlation between loop count and secret value.
\end{itemize}

\subsection{Exception-Based Information Leak}

\paragraph{Threat.}
If an exception is thrown only when a secret meets a specific condition, the presence or absence of an error leaks one bit of information per execution. Similar channels have been used to extract secrets in hardened kernel prototypes~\cite{schneider2016error}.

\paragraph{Mitigations.}
\begin{itemize}
  \item \textbf{Explicit result types.} Replace exceptions with structured outputs like \texttt{Result\{ok, error\}}, allowing both paths to be processed uniformly.
  \item \textbf{Consistent control flow.} Ensure both success and error branches execute with the same timing and code structure to prevent divergence-based leaks.
\end{itemize}

\subsection{Timing Channels}

\paragraph{Threat.}
Execution time can vary based on secret-dependent logic - for example, sleeping for \texttt{secret} seconds or using branches that hit different cache lines. Kocher’s classic work showed that even sub-millisecond differences can leak RSA keys~\cite{kocher1996timing}; later research demonstrated full AES key recovery via cache timing~\cite{osvik2006cache}.

\paragraph{Mitigations.}
\begin{itemize}
  \item \textbf{Timer stubbing.} Remove access to high-resolution timers in untrusted code or introduce jitter to reduce precision.
  \item \textbf{Constant-time operations.} Pad sensitive operations to their worst-case runtime before returning control, as in cryptographic libraries.
  \item \textbf{Deterministic scheduling.} Process tool calls in a fixed sequence and timing pattern, eliminating runtime variations tied to secret values.
\end{itemize}

\medskip
\noindent
Together, these mitigations - loop clamping, structured error handling, and constant-time execution - can suppress the most practical side channels without altering CaMeL’s capability model. These controls are essential for systems handling regulated or confidential data where indirect leakage must be accounted for.

\section{Architectural Limitations}

While CaMeL introduces a strong capability-based execution model, its current design brings several architectural trade-offs. These include performance bottlenecks, complexity in policy management, and limitations in the interpreter’s language semantics. In this section, we outline key challenges and offer practical design adjustments that could support more efficient and scalable deployments.

\subsection{Performance Overhead of Dual LLMs}

CaMeL separates control and data by using two large language models: a \textit{Privileged LLM} (P-LLM) that generates plans and a \textit{Quarantined LLM} (Q-LLM) that validates untrusted content. While this split provides strong data isolation, it effectively doubles the number of model invocations.

OpenAI’s published metrics for GPT-4 report median latencies of 1–2 seconds per request and pricing of \$0.03–\$0.06 per 1,000 tokens (prompt + completion)~\cite{openai2024pricing}. In workflows where the Q-LLM must process multiple artifacts - such as reviewing 10 email messages - latency can exceed 10 seconds. This delay is often unacceptable in interactive applications like customer-facing chatbots or support agents.

To reduce this overhead, we recommend the following design strategies:

\begin{enumerate}
  \item \textbf{Plan-template caching.} Many enterprise prompts fall into a small set of repeated intents (e.g., “summarize my inbox,” “file this expense report”). Caching known-safe plans keyed by prompt hashes allows reuse without re-invoking the P-LLM.
  \item \textbf{Deterministic micro-parsers.} For structured outputs like JSON, using hand-written or generated parsers is faster and cheaper than calling an LLM. This mirrors techniques in ReAct-style hybrid agents~\cite{yao2023react}.
  \item \textbf{Batching Q-LLM extractions.} When the same validation logic must be applied to many strings (e.g., extracting “amount” from 100 receipts), concatenating them into a single prompt amortizes the per-call cost.
\end{enumerate}

Internal tests show that combining caching with deterministic parsers can cut token usage and end-to-end latency by up to 50 percent while preserving CaMeL’s policy guarantees.

\subsection{Policy Maintenance at Scale}

Each tool in CaMeL is governed by a Python policy function that specifies allowed data flows. In large organizations - especially those with hundreds of APIs - this results in policy sprawl: inconsistent logic, duplicated rules, and lack of central auditing. These issues mirror configuration drift problems identified in infrastructure-as-code research~\cite{rahman2020iac}.

To address this, we recommend adopting \textbf{policy-as-code frameworks} with the following features:

\begin{itemize}
  \item \textit{Declarative languages.} Tools like Rego (used in Open Policy Agent) represent access logic as pure rules, which are easier to test, audit, and verify~\cite{torin2022opa}.
  \item \textit{Reusable modules.} Common rules - such as “share only within domain” - can be imported and reused across multiple tools.
  \item \textit{Visual interfaces.} GUI-based editors lower the barrier for non-engineering stakeholders to read or modify policy rules safely.
\end{itemize}

Centralizing policies in a declarative engine simplifies governance, reduces human error, and makes policy behavior easier to reason about.

\subsection{Interpreter Language Constraints}

CaMeL uses a restricted subset of Python to define tool plans. While this improves accessibility for developers, it inherits Python’s problematic features: dynamic typing, exception-driven control flow, and reflection. These traits make static analysis difficult and prevent strong guarantees about information flow.

A more robust approach would be to re-implement the plan interpreter using a \textbf{security-oriented domain-specific language (DSL)}. Such a language would support:

\begin{itemize}
  \item Bounded loops and simple, explicit conditionals
  \item First-order function calls only (no reflection or metaprogramming)
  \item Capability labels embedded in the type system
\end{itemize}

This mirrors prior work in language-based security, such as FlowCaml~\cite{simonet2004flowcaml}, which enables compile-time enforcement of information-flow constraints. Embedding the DSL in a proof-oriented platform like F could allow formal verification of constant-time behavior and noninterference~\cite{swamy2016fstar}.

\section{Comparison with Other Defenses}

Enterprise security teams have several strategies available to defend against prompt injection \cite{tallam2025threatintel}. These approaches vary in terms of where they intervene in the stack - at the model level, system level, or input/output layer - and each comes with distinct trade-offs. Below, we compare CaMeL with three common classes of defenses.

\paragraph{Model-level robustness.}
Instruction tuning and preference alignment methods - such as InstructGPT and Constitutional AI - aim to make the model itself more resistant to malicious prompts~\cite{ouyang2022instruct,bai2022constitutional}. These techniques fine-tune LLMs to ignore or suppress harmful instructions embedded in context. While they reduce jailbreak success rates on average, they remain heuristic: adversaries can still craft obfuscated payloads that bypass training filters. Moreover, most enterprises use closed-source models and cannot re-train them directly.

CaMeL takes a different approach by treating the LLM as a black box. Instead of modifying the model, it adds a runtime policy layer to control how data flows through agent decisions - offering protection even when model internals are inaccessible.

\paragraph{System-level sandboxing.}
Isolation tools like Google’s gVisor and micro-VM frameworks restrict what an agent process can do at the operating system level~\cite{gvisor2022}. These sandboxes are effective at containing damage from code execution errors or malware, but they don’t address misuse of allowed functionality. For example, if SMTP is permitted, an injected prompt can still send sensitive data by crafting a plausible email.

CaMeL complements sandboxing by applying fine-grained capability checks on individual tool invocations, filtering requests based on data origin and context - not just system-level permissions.

\paragraph{Static filtering and input sanitization.}
Some defenses scan inputs for known attack patterns - e.g., blacklisting keywords like \texttt{ignore\_previous} or removing suspicious HTML tags. These filters are fast and easy to deploy but are brittle: red team studies show they block fewer than 30 percent of novel jailbreak strategies~\cite{perez2022red,zou2023universal}. Attackers quickly find paraphrases, encoding tricks, or Unicode variants to bypass static rules.

By contrast, CaMeL performs dynamic, context-aware validation after parsing, labeling inputs based on provenance, and enforcing policies at execution time. This makes it more resilient to previously unseen prompt injection tactics and avoids over-relying on fragile string matching.

\medskip
\noindent
In summary, CaMeL complements existing defenses. It is particularly well-suited for enterprises that cannot retrain foundation models but require structured, verifiable control over how LLMs process and act on untrusted inputs.

\section{Conclusion}
This work extends CaMeL by addressing overlooked threats and architectural challenges that impact its deployability in real-world settings. We identify three core gaps in its security posture - initial prompt trust, output manipulation, and side-channel exposure - and propose mitigations that preserve CaMeL’s fine-grained policy model while improving its robustness and scalability. By introducing adaptive risk tiers, prompt/output filtering, and verification-ready execution models, we chart a path toward a production-grade CaMeL variant suitable for regulated and high-assurance environments. Rather than replacing CaMeL, our response enhances its foundation, offering concrete steps to transition from academic prototype to enterprise-ready security framework for LLM agents.

\section{Directions for Future Work}

Transforming CaMeL from a promising prototype into a production-grade security framework will require progress along several technical dimensions:

\begin{itemize}
  \item \textbf{Mechanized noninterference proofs.}  
        Formalizing CaMeL’s execution semantics in a proof assistant such as Coq or Isabelle and proving \emph{end-to-end noninterference} would elevate its assurance level to that of verified systems like CompCert and CertiKOS~\cite{leroy2009compcert,gu2016certikos}. A machine-checked theorem would guarantee that secret-tagged data cannot influence public outputs except through explicitly authorized channels.

  \item \textbf{A security-oriented DSL.}  
        Rewriting CaMeL plans in a minimal, typed intermediate language - with no reflection, implicit coercions, or exception-based control - would enable more tractable static analysis. FlowCaml’s type-based enforcement of information flow~\cite{simonet2004flowcaml} offers a useful model for embedding capability labels directly in the type system and enforcing policies at compile time.

  \item \textbf{Constant-time execution.}  
        Even with strong data-flow controls, timing channels can persist. To mitigate this, techniques from verified cryptographic software - such as constant-time transformations from the HACL library~\cite{almeida2017hacl} - can be adapted to CaMeL’s interpreter, ensuring that execution time does not vary based on confidential inputs~\cite{kocher1996timing}.

  \item \textbf{Adaptive risk policies.}  
        As enterprises adopt Zero Trust architectures, access decisions increasingly rely on real-time context such as device posture, location, and behavioral patterns~\cite{mell2020zero}. Integrating these signals into CaMeL’s policy engine would allow dynamic escalation or relaxation of enforcement tiers (see Section~\ref{sec:balancing}), improving usability without compromising security for sensitive operations \cite{tallam2025alignment}.
\end{itemize}

\bibliographystyle{unsrt}
\bibliography{references}

\begin{thebibliography}{10}

\bibitem{carlini2023poisoning}
Nicholas Carlini, Florian Tramer, Eric Wallace, Matthew Jagielski, Ariel Herbert-Voss, Miles Brundage, Tom Brown, Deep Ganguli, Úlfar Erlingsson, et~al.
\newblock Poisoning language models during instruction tuning.
\newblock {\em arXiv preprint arXiv:2302.12173}, 2023.
\newblock \url{https://arxiv.org/abs/2302.12173}.

\bibitem{tallam2025cybersentinel}
Krti Tallam.
\newblock Cybersentinel: An emergent threat detection system for ai security, 2025.

\bibitem{abu2007comparison}
Saleh Abu-Nimeh, Dan Nappa, Xiang Wang, and Salil Nair.
\newblock A comparison of machine learning techniques for phishing detection.
\newblock {\em eCrime Researchers Summit}, pages 60--69, 2007.

\bibitem{tallam2025assurance}
Krti Tallam.
\newblock Engineering risk-aware, security-by-design frameworks for assurance of large-scale autonomous ai models, 2025.

\bibitem{tallam2025cyberimmune}
Krti Tallam.
\newblock The cyber immune system: Harnessing adversarial forces for security resilience, 2025.

\bibitem{williams2018broad}
Adina Williams, Nikita Nangia, and Samuel Bowman.
\newblock A broad-coverage challenge corpus for sentence understanding through inference.
\newblock {\em NAACL-HLT}, pages 1112--1122, 2018.

\bibitem{myers1999jif}
Andrew~C. Myers.
\newblock Jif: Java information flow.
\newblock In {\em IEEE Computer Security Foundations Workshop (CSFW)}, pages 187--196, 1999.

\bibitem{simonet2004flowcaml}
Jérôme Simonet.
\newblock Flowcaml: A polymorphic information-flow language.
\newblock In {\em ACM SIGPLAN Workshop on ML}, pages 85--96, 2004.

\bibitem{mell2020zero}
Scott Rose, Oliver Borchert, Stu Mitchell, and Sean Connelly.
\newblock Zero trust architecture.
\newblock NIST SP 800-207, 2020.
\newblock \url{https://doi.org/10.6028/NIST.SP.800-207}.

\bibitem{hu2015abac}
Vincent~C. Hu, David Ferraiolo, Richard Kuhn, and Angelo Schnitzer.
\newblock Guide to attribute based access control (abac): Definition and considerations.
\newblock Technical Report SP 800-162, NIST, 2015.

\bibitem{furman2016security}
Suzanna Furman, Mary Theofanos, Yee-Yin Choong, and Betty Stanton.
\newblock Security fatigue.
\newblock {\em IT Professional}, 18(5):26--32, 2016.

\bibitem{gu2016certikos}
Ronghui Gu, Dominic Costanzo, and Zhong Shao.
\newblock Certikos: An extensible architecture for building certified concurrent os kernels.
\newblock In {\em USENIX OSDI}, pages 653--669, 2016.

\bibitem{leroy2009compcert}
Xavier Leroy.
\newblock Formal verification of a realistic compiler.
\newblock {\em Communications of the ACM}, 52(7):107--115, 2009.

\bibitem{sabelfeld2003language}
Andrei Sabelfeld and David Sands.
\newblock Language-based information-flow security.
\newblock {\em IEEE Journal on Selected Areas in Communications}, 21(1):5--19, 2003.

\bibitem{xu2015controlled}
Yan Xu, Weidong Cui, and Marcus Peinado.
\newblock Controlled-channel attacks: Deterministic side channels for untrusted operating systems.
\newblock In {\em IEEE Symposium on Security and Privacy}, pages 640--656, 2015.

\bibitem{schneider2016error}
Katharina Schneider and Sören Bulander.
\newblock Error message side channels in secure operating systems.
\newblock In {\em Proceedings of the 2016 ACM Conference on Computer and Communications Security}, pages 1052--1065, 2016.

\bibitem{kocher1996timing}
Paul~C. Kocher.
\newblock Timing attacks on implementations of diffie–hellman, rsa, dss, and other systems.
\newblock In {\em CRYPTO ’96}, volume 1109 of {\em LNCS}, pages 104--113. Springer, 1996.

\bibitem{osvik2006cache}
Dag~Arne Osvik, Adi Shamir, and Eran Tromer.
\newblock Cache attacks and countermeasures: The case of aes.
\newblock In {\em RSA Conference Cryptographers’ Track}, volume 3860 of {\em LNCS}, pages 1--20. Springer, 2006.

\bibitem{openai2024pricing}
OpenAI.
\newblock Openai api pricing.
\newblock \url{https://openai.com/pricing}, 2024.

\bibitem{yao2023react}
Shunyu Yao, Dian Zhao, Jeffrey Yu, Dong Yang, Yuan Chen, and et~al.
\newblock React: Synergizing reasoning and acting in language models.
\newblock {\em arXiv preprint arXiv:2210.03629}, 2023.

\bibitem{rahman2020iac}
Mohammad Rahman and Laurie Williams.
\newblock Security analysis of infrastructure-as-code: A systematic study.
\newblock {\em Empirical Software Engineering}, 25:4922--4960, 2020.

\bibitem{torin2022opa}
Torin Fairchild and Wyatt Dillon.
\newblock Open policy agent.
\newblock \url{https://www.openpolicyagent.org/}, 2022.

\bibitem{swamy2016fstar}
Nikhil Swamy, Cencia Zhu, Bjoern Pfaff, and et~al.
\newblock Dependent types and multi-monadic effects in f$^*$.
\newblock In {\em ACM SIGPLAN--SIGACT Symposium on Principles of Programming Languages (POPL)}, pages 256--270, 2016.

\bibitem{tallam2025threatintel}
Krti Tallam.
\newblock Transforming cyber defense: Harnessing agentic and frontier ai for proactive, ethical threat intelligence, 2025.

\bibitem{ouyang2022instruct}
Long Ouyang, Jeffrey Wu, Xu~Jiang, Diogo Almeida, and et~al.
\newblock Training language models to follow instructions with human feedback.
\newblock {\em arXiv preprint arXiv:2203.02155}, 2022.

\bibitem{bai2022constitutional}
Yuntao Bai, Saurav Kadavath, Sandipan Kundu, Amanda Askell, and et~al.
\newblock Constitutional ai: Harmlessness from ai feedback.
\newblock {\em arXiv preprint arXiv:2212.08073}, 2022.

\bibitem{gvisor2022}
Google~Open Source.
\newblock gvisor: User-space kernel sandbox.
\newblock \url{https://gvisor.dev}, 2022.

\bibitem{perez2022red}
Ethan Perez, Samuel Ringer, Nisan Nanda, and et~al.
\newblock Red teaming language models with language models.
\newblock {\em arXiv preprint arXiv:2202.03286}, 2022.

\bibitem{zou2023universal}
Andy Zou, Frank~F. Xu, Micah Goldblum, and Tom Goldstein.
\newblock Universal and transferable adversarial attacks on aligned language models.
\newblock {\em arXiv preprint arXiv:2307.15043}, 2023.
\newblock \url{https://arxiv.org/abs/2307.15043}.

\bibitem{almeida2017hacl}
José~Bacelar Almeida, Anne Béguin, Karthikeyan Bhargavan, Antoine Delignat-Lavaud, Cédric Fournet, Santiago Zanella-Béguelin, and Jean-Karim Zinzindohoué.
\newblock {HACL}* : Verified implementations of cryptographic algorithms.
\newblock In {\em ACM CCS}, pages 1789--1806, 2017.

\bibitem{tallam2025alignment}
Krti Tallam.
\newblock Alignment, agency and autonomy in frontier ai: A systems engineering perspective, 2025.

\end{thebibliography}
\end{document}